\documentclass{aa}

\usepackage[varg]{txfonts}
\usepackage{graphicx}
\usepackage{amsmath}
\usepackage[colorlinks=true,linkcolor=magenta,citecolor=blue,filecolor=cyan,urlcolor=purple,final=true]{hyperref}

\makeatletter
\renewcommand*\aa@pageof{, page \thepage{} of \pageref*{LastPage}}
\makeatother
 
\begin{document} 

   \title{Extended radio emission associated with a breakout eruption from the back side of the Sun}
        \titlerunning{Radio emission following a back side eruption}

   \author{D.~E.~Morosan \inst{1}
        \and
        E.~Palmerio \inst{1} 
        \and
        B.~J.~Lynch \inst{2}
        \and
          E.~K.~J.~Kilpua \inst{1} 
          }

   \institute{Department of Physics, University of Helsinki, P.O. Box 64, FI-00014, Helsinki, Finland \\
              \email{diana.morosan@helsinki.fi}
           \and
           Space Sciences Laboratory, University of California--Berkeley, Berkeley, CA 94720, USA
             }

   \date{Received ; accepted }

 
  \abstract
    {Coronal mass ejections (CMEs) on the Sun are the largest explosions in the Solar System that can drive powerful plasma shocks. The eruptions, shocks, and other processes associated to CMEs are efficient particle accelerators and the accelerated electrons
in particular can produce radio bursts through the plasma emission mechanism.}
    {Coronal mass ejections and associated radio bursts have been well studied in cases where the CME originates close to the solar limb or within the frontside disc. Here, we study the radio emission associated with a  CME eruption on the back side of the Sun on 22 July 2012.}
    {Using radio imaging from the Nan{\c c}ay Radioheliograph, spectroscopic data from the Nan{\c c}ay Decametric Array, and extreme-ultraviolet observations from the Solar Dynamics Observatory and Solar Terrestrial Relations Observatory spacecraft, we determine the nature of the observed radio emission as well as the location and propagation of the CME.}
    {We show that the observed low-intensity radio emission corresponds to a type II radio burst or a short-duration type IV radio burst associated with a CME eruption due to breakout reconnection on the back side of the Sun, as suggested by the pre-eruptive magnetic field configuration. The radio emission consists of a large, extended structure, initially located ahead of the CME, that corresponds to various electron acceleration locations.}
    {The observations presented here are consistent with the breakout model of CME eruptions. The extended radio emission coincides with the location of the current sheet and quasi-separatrix boundary of the CME flux and the overlying helmet streamer and also with that of a large shock expected to form ahead of the CME in this configuration.}


\maketitle


\section{Introduction} \label{sec:intro}

{Coronal mass ejections (CMEs) are large eruptions of magnetised plasma that are regularly ejected from the Sun. These phenomena are generally due to twisted or sheared magnetic fields in the corona that eventually become unstable. Currently, there are two main scenarios that lead to the formation of a CME: sheared arcades that can become unstable due to reconnection in either a breakout configuration \citep{an99,ly08} or via tether cutting \citep{mo01} and flux cancellation \citep{li03}, or ideal instabilities acting on pre-eruptive flux rope configurations such as the kink or torus instabilities  \citep{to05, kl06}. Each of these CME initiation mechanisms results in flare reconnection below the erupting field and plasma structure and produces a flux rope ejecta usually seen in white light as a bright core and a cavity, surrounded by a bright compression front \citep{ch97, au10, vo13}. }

{Coronal mass ejections can be powerful drivers of plasma shocks that accelerate electrons to high energies. These electrons in turn generate bursts of radiation at metre and decimetre wavelengths through the plasma emission mechanism, such as type II radio bursts \citep{kl02} and herringbones \citep{ro59, ca87}. Type II radio bursts are characterised by emission bands in dynamic spectra, with a frequency ratio of 2:1 representing emission at the fundamental and harmonic of the plasma frequency \citep{ne85}. Herringbone bursts are characterised by `bursty' signatures superimposed on type II bursts or occurring on their own, and are signatures of individual electron beams accelerated at the CME shock \citep{mo19a}. Coronal mass ejections can also be accompanied by continuum emission at decimetric and metric wavelengths, namely type IV radio bursts that can show stationary or moving sources, or both, emitted by various emission mechanisms \citep{ba98, mo19b}. In this case, the electrons are believed to originate following magnetic reconnection processes \citep{de12}. So far, radio bursts have been widely studied in plane-of-sky observations where close associations have been found between type II radio bursts and CME shocks \citep{sm70, zu18}. However, the plane-of-sky locations of radio bursts are affected by projections, and accurate estimates of the location of this emission in relation to the CME rely on observations of eruptions close to the solar limb.}

   \begin{figure*}[ht]
   \centering
          \includegraphics[width=0.8\textwidth]{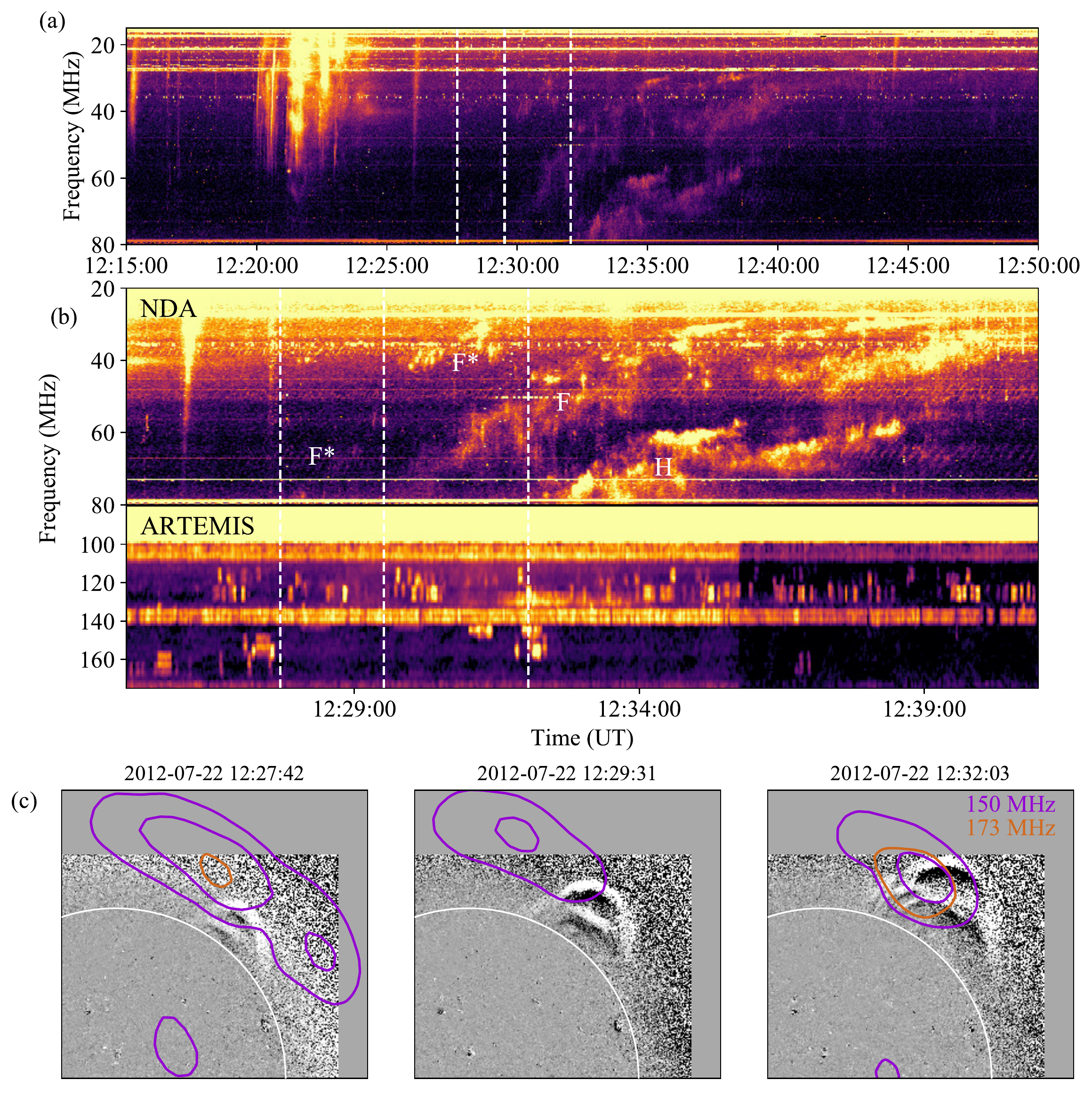}
      \caption{Dynamic spectra and images of the radio bursts and CME observed on 22 July 2012. (a) Dynamic spectrum from the NDA showing a faint type II radio burst occurring a few minutes after a set of type III bursts. (b) Zoom-in of the dynamic spectrum in (a) showing the type II burst in more detail as observed by the NDA at 20--80~MHz and the higher frequency observations from the ARTEMIS-IV Radiospectrograph \citep{ko06} at 80--175~MHz that did not observe any type II signatures or other distinguishable radio bursts. (c) Contours of the radio emission observed in images from the NRH at 150 (purple) and 173 (brown) MHz overlaid on SDO/AIA 211~{\AA} running-difference images of the Sun. }
         \label{fig1}
   \end{figure*}

{In this paper, we present observations of radio emission associated with a CME that erupted on 22 July 2012 from the back side of the visible solar disc. These unique observations are analysed in conjunction with the Solar Terrestrial Relations Observatory \citep[STEREO;][]{ka08} twin spacecraft, which provide two new perspectives of the CME eruption and have allowed reconstruction of the 3D spatial location of the radio emission. In Sect.~\ref{sec:analysis}, we give an overview of the observations and data analysis techniques used. In Sect.~\ref{sec:results}, we present the results of the analysis of the CME eruption and associated radio emission, as well as the pre-eruptive configuration, which are further discussed in Sect.~\ref{sec:conclusions}.}


   \begin{figure*}[ht]
   \centering
          \includegraphics[width=0.9\textwidth]{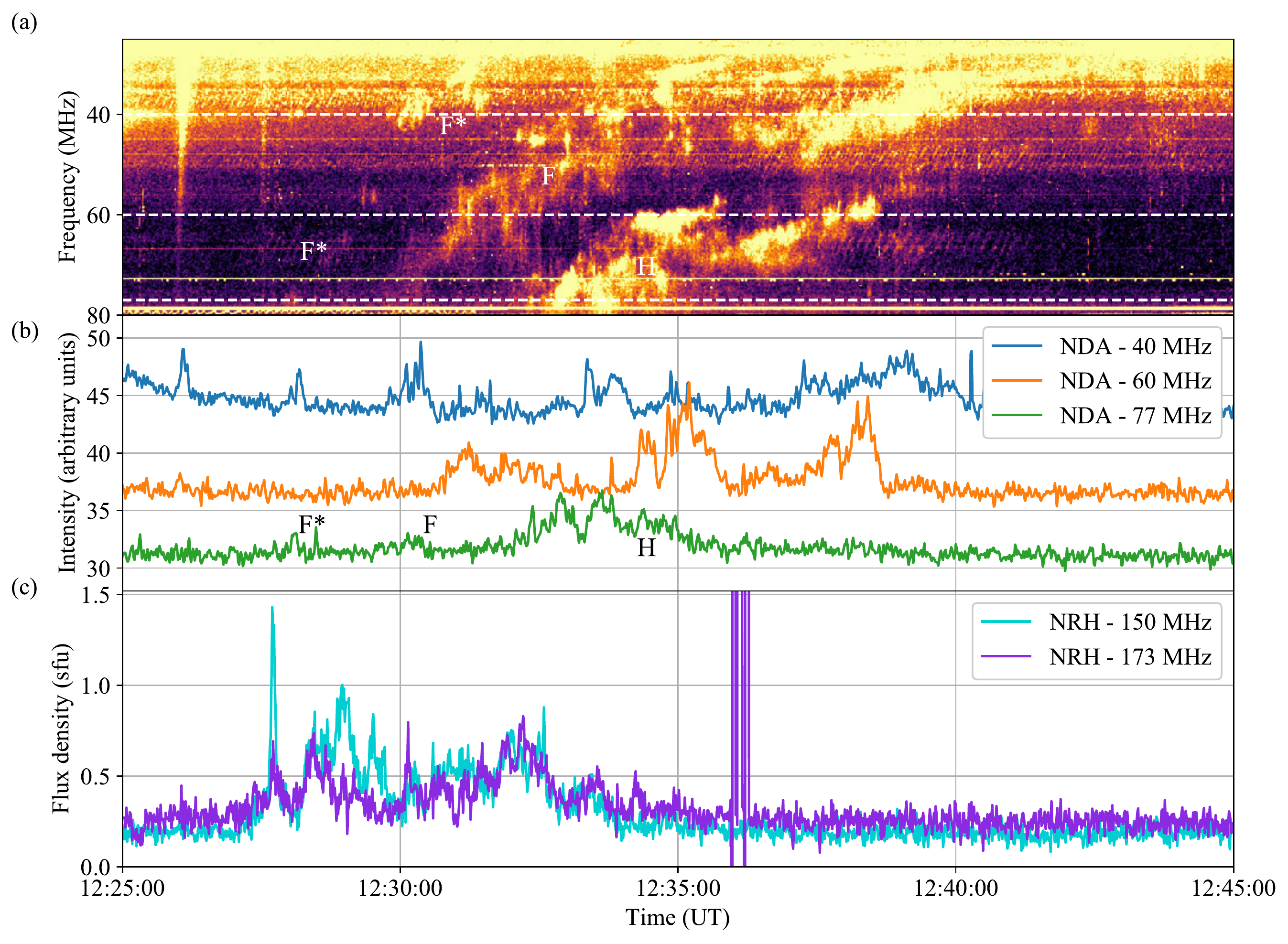}
      \caption{Dynamic spectra and images of the type II radio burst and flux comparison of the type II and radio sources observed by NRH. (a) Dynamic spectrum from the NDA showing the type II radio burst. (b) Intensity of the type II burst at three frequencies: 40, 60, and 77~MHz, along the dashed white lines in (a). (c) Flux density of the radio sources observed by the NRH during the same time period as in (a) at 150 and 173~MHz.}
         \label{fig2}
   \end{figure*}

\section{Observations and data analysis} \label{sec:analysis}

{On 22 July 2012, a CME occurred on the back side of the visible solar disc, associated with a faint type II radio burst observed by the Nan{\c c}ay Decametric Array \citep[NDA;][]{bo80} in the 10--80~MHz frequency range (Fig.~\ref{fig1}). The type II burst started at ${\sim}$12:30~UT in the NDA data and lasted for ${\sim}15$~minutes. The type II burst was faint compared to preceding activity in the form of type III radio bursts. Multiple emission lanes were observed, including two main emission bands that represent fundamental and harmonic emission (denoted F and H in Fig.~\ref{fig1}b, which is a zoomed-in image from Fig.~\ref{fig1}a). The F--H emission bands show band splitting, which is believed to occur due to plasma emission upstream and downstream of the CME shock \citep{vr01,vr02}. Fainter emission lanes were also observed before the F--H bands, denoted F* in Fig.~\ref{fig1}b. The type II radio burst is likely associated with a  CME that is behind the limb (Fig.~\ref{fig1}c), since no eruptive signatures \citep[e.g.,][]{hu01} were observed on the visible solar disc based on images by the Atmospheric Imaging Assembly \citep[AIA;][]{le12} onboard the Solar Dynamics Observatory \citep[SDO;][]{pe12}. }

{The radio sources associated with the CME and observed with the Nan{\c c}ay Radioheliograph \citep[NRH;][]{ke97} are shown in Fig.~\ref{fig1}c, where the radio source contours at 150 and 173~MHz are overlaid on three different images of the Sun taken by SDO/AIA. The radio contours represent the 30\% and 70\% of the maximum intensity levels in each image. The overall radio source contours in Fig.~\ref{fig1}c are large compared to the CME extent in its early eruption phase observed in the plane-of-sky view from SDO. Such large radio sources are sometimes observed in the case of type IV bursts \citep{ma03, ca17, mo19b} and have rarely been observed before in the case of type II bursts \citep{sm70}. The NRH radio sources were only observed at 150 and 173~MHz and consist of bursty emission that is unpolarised with a low flux density (see Movie~1 accompanying this paper). Unfortunately, the onset of the type II harmonic band at higher frequencies was not seen in any of the other available data sets (see e.g., Fig.~\ref{fig1}b). }


\section{Results} \label{sec:results}

\subsection{The nature of the radio emission}

{Further insight into the nature of the NRH radio sources can be obtained by investigating their flux density evolution through time. We extract the intensity of the type II radio burst observed by the NDA to compare it to the flux of the NRH sources. The dynamic spectrum and intensity of the type II burst are shown in Fig.~\ref{fig2}a--b. The intensity of this emission is shown at three separate frequencies, namely 40, 60, and 77~MHz, along the dashed white lines reported in Fig.~\ref{fig2}a. For comparison, we estimated the NRH flux density of the observed radio sources at 150 and 173~MHz (Fig.~\ref{fig2}c). Although both the type II emission lanes and NRH sources are bursty in nature and short in duration, there is no clear correlation between the two emissions. There are however bursts observed with the NRH that are co-temporal with the F and F* components. The bursty nature, short bandwidth and short duration indicate that the NRH radio sources are emitted by the plasma emission mechanism. Thus, there are two possibilities to explain the radio emission observed by the NRH: either some of this emission represents the harmonic lane of the type II fundamental observed at lower frequencies (Fig.~\ref{fig1}b) or this extended emission corresponds to a short-duration type IV radio burst. These two scenarios are investigated further below.} 

{If the radio source observed with the NRH at higher frequency corresponds to a type II burst, then some of the radio contours in Fig.~\ref{fig1}c represent the harmonic emission lanes of the type II radio burst observed in the NDA dynamic spectrum (Fig.~\ref{fig1}b). Both fundamental emission lanes, labelled F and F* in Fig.~\ref{fig1}b, start at a frequency of 80~MHz or higher that would correspond to harmonic emission at the same time at NRH frequencies. Therefore, the radio contours at 12:27:42 and 12:29:31~UT in Fig.~\ref{fig1}c occur at the same time and frequency as the possible type II harmonic lanes inferred from the fundamental lanes at lower frequencies. If some of this emission corresponds to a type II burst, then these radio sources indicate that the type II electrons are accelerated by a shock ahead of the CME front.}

   \begin{figure*}[ht]
   \centering
          \includegraphics[width=0.98\textwidth]{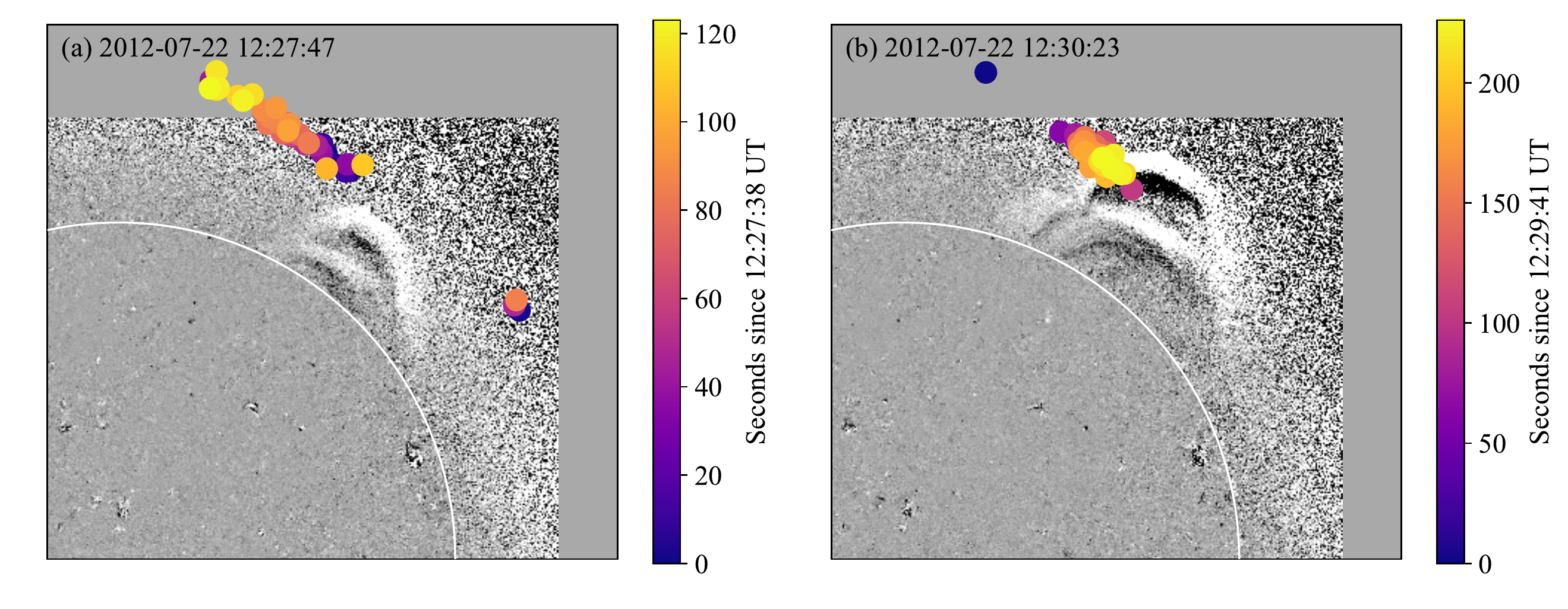}
      \caption{Centroids of radio sources overlaid on running-difference images from SDO/AIA at 211~{\AA}. The centroids have been separated into two
plane-of-sky components: (a) an earlier moving component starting at 12:27:38~UT and a (b) later stationary component starting at 12:29:41~UT.}
         \label{fig3}
   \end{figure*}

{The radio sources observed with the NRH could also correspond to a short-duration type IV radio burst, since type IV bursts can sometimes show such extended structures \citep{ma03, ca17, mo19b}. In this case the radio emission would not originate from a CME shock but would likely be due to reconnection during the eruption process. Type IV bursts can also result from the gyro-synchrotron motion of electrons that are trapped within the CME magnetic field, but due to the bursty and narrowband nature of the emission, its relatively short duration, and the positioning of the radio source relative to the CME, we do not deem this to be a likely scenario. Since the radio emission is initially observed ahead of the CME, as seen from the plane of sky view from SDO, such reconnection processes must also occur ahead of the CME front.}

{The radio source contours appear to change in extent and location throughout the duration of the radio emission (for the full evolution of the CME and associated radio emission as seen from Earth, see Movie~2 accompanying this paper). We extract the centroids of the most intense radio source observed in each image. We find two components of the radio emission: an earlier moving component from 12:27:38~UT that lasts for ${\sim}2$~minutes and is part of the extended structure in Fig.~\ref{fig1}c (Fig.~\ref{fig3}a), and a later stationary component starting at 12:29:41~UT that coincides with a single radio source, smaller in extent in Fig.~\ref{fig1}c, and located at the eastern flank of the CME in the plane-of-sky view (Fig.~\ref{fig3}b). Since we are tracking the maximum emission source, the centroids marked in Fig.~\ref{fig3}a could also correspond to individual radio bursts becoming brighter at farther locations on the eastern limb and not necessarily the movement of the same radio burst. However, the radio emission intensifies in the direction of the CME expansion towards the north pole. The moving component could correspond to the type II radio burst observed at NDA frequencies, since it shows a propagation path similar to that of the CME, observed before with type II bursts and herringbones \citep{ca13, mo19a}. However, moving type IV radio bursts can also show a propagation path in the direction of the CME expansion, inside or outside CMEs, while the stationary component resembles a stationary type IV burst on top of the eruption site. This has been observed before in cases where type IV bursts show both moving and stationary components \citep{mo19b}. Therefore, the emission observed with the NRH could be either a type II or a type IV burst, or both.}

   \begin{figure*}[ht]
   \centering
          \includegraphics[width=0.95\linewidth]{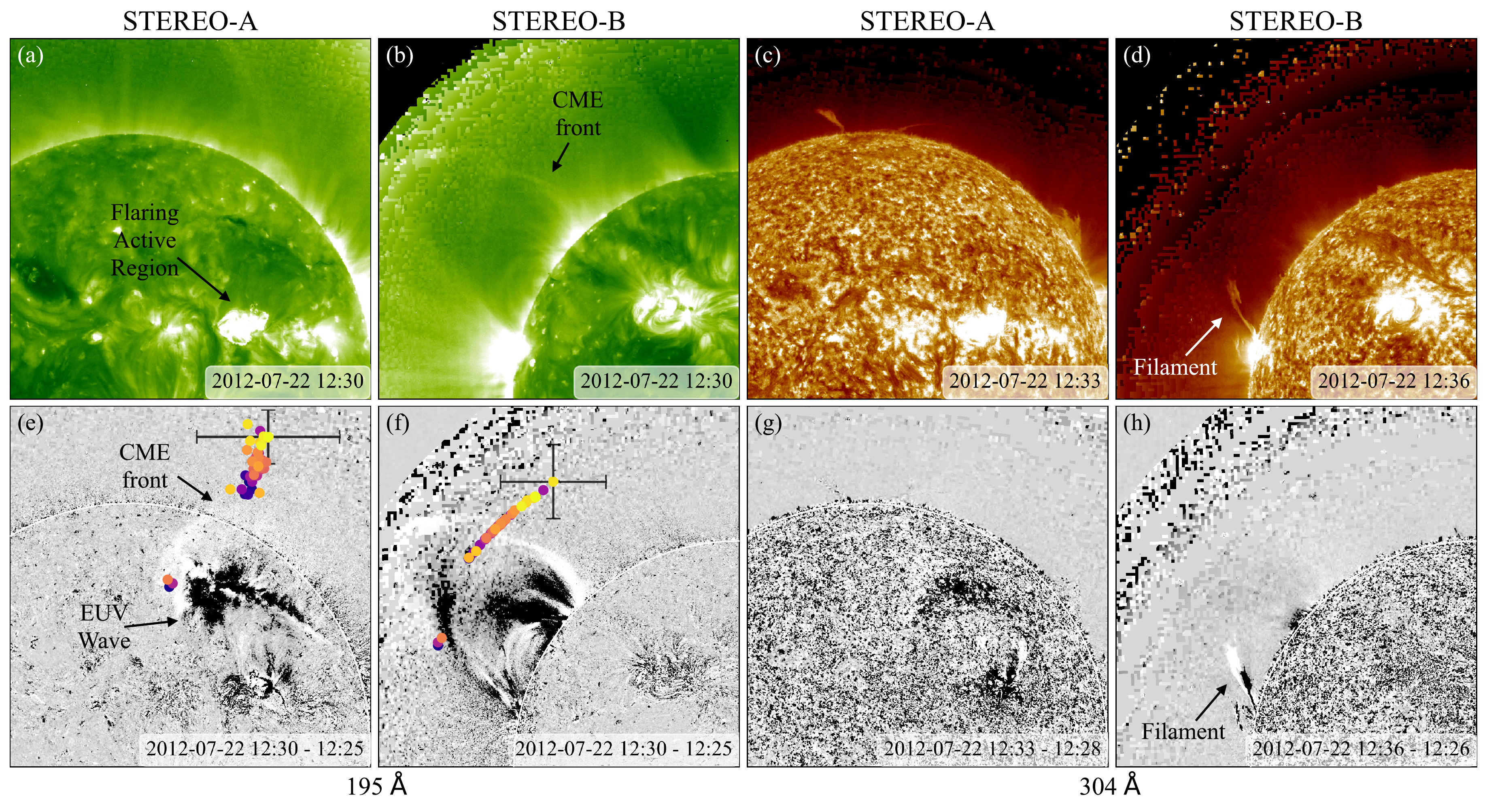}
      \caption{Coronal mass ejection eruption and radio emission from the STEREO perspectives. (a--d) Intensity and (e--h) running-difference images taken by STEREO/SECCHI/EUVI-A and -B at the 195~{\AA} (left panels) and 304~{\AA} (right panels) channels around the time of the CME eruption. The radio centroids (together with a $\pm0.3~R_\odot$ error shown for one example centroid) have been projected onto the STEREO-A (e) and -B (f) views. The colour-bar for the radio centroids is the same as in Fig.~\ref{fig3}a.}
         \label{fig4}
   \end{figure*}

\subsection{The CME origin and location of the associated radio emission}

{In order to confirm that the 22 July 2012 CME originated from the back side of the solar disc, we investigate images from the Sun Earth Connection Coronal and Heliospheric Investigation \citep[SECCHI;][]{ho08} suite onboard the twin STEREO spacecraft, which were located ${\sim}120^{\circ}$W (STEREO-A) and ${\sim}115^{\circ}$E (STEREO-B) away from Earth at that time. We find low-coronal signatures of the CME on-disc in STEREO-A data and off-limb in STEREO-B data. Images of the eruption from the Extreme-Ultraviolet Imager (EUVI) onboard STEREO are shown in Fig.~\ref{fig4} (for the full evolution of the CME from the STEREO viewpoints, see Movie~3 accompanying this paper). The eruptive event commenced with a small active-region flare and was characterised by the presence of an extreme ultraviolet (EUV) wave and the ejection of a filament. The resulting CME had a radial speed of ${\sim}740$~km/s in the low corona from the STEREO-B perspective, where projection effects are expected to be minimal.}

{An approximate location of the radio bursts position can be obtained by projecting the moving radio centroids in Fig.~\ref{fig3}a onto the STEREO perspectives. In order to achieve this, we obtain an estimate of the radial height of the radio bursts in the SDO plane using a radial electron density model to relate the frequency of plasma radiation to coronal heights. We use a Newkirk four-fold density model \citep{new61} and assume that the radio bursts are emitted at the harmonic plasma emission, since this back-side emission is unpolarised \citep{du82} and must occur at a relatively high altitude behind the limb to be observed outside the solar limb in the SDO plane. The approximate height that we obtain is $1.64~R{_\odot}$ from the centre at a frequency of 150~MHz. To this, we have added an uncertainty of $\pm0.3~R_\odot$ that reflects a range of possible heights based on models of the background solar corona \citep[e.g.][]{sa77} and models of a more active solar corona (e.g.\ the Newkirk 6-fold model). From this height we obtain the $z$-coordinate of the centroids, which is assumed negative since the emission originated behind the solar disc. The $x$,$y$,$z$-coordinates of the centroids are then projected onto the STEREO-A (Fig.~\ref{fig4}e) and STEREO-B (Fig.~\ref{fig4}f) images. Since we used a density model to obtain the 3D coordinates of the radio centroids, their projected position on STEREO is not exact, however the propagation direction and location relative to the CME front do not change as reflected by the range in positions within the extent of the error bars in Fig.~\ref{fig4}e--f. }

{In STEREO-A, the centroids are located close to the apex of the eruption and propagate outwards in the same direction as the CME (Fig.~\ref{fig4}e). In STEREO-B, the majority of centroids are located on the western CME flank, propagating towards the north pole in the same direction as the CME lateral and non-radial propagation expansion (Fig.~\ref{fig4}f). Although the error bars for the de-projected centroids may appear large, in STEREO-A the centroids are still located close to the CME apex when taking the uncertainty in position into account. However, in STEREO-B the centroids could also be located closer to the CME core. Based on the STEREO-A and SDO perspectives, the centroids in STEREO-B are also most likely located outside the CME but they are affected by projection effects.}

\subsection{The pre-eruptive coronal magnetic field configuration}

   \begin{figure*}[ht]
   \centering
          \includegraphics[width=0.95\linewidth]{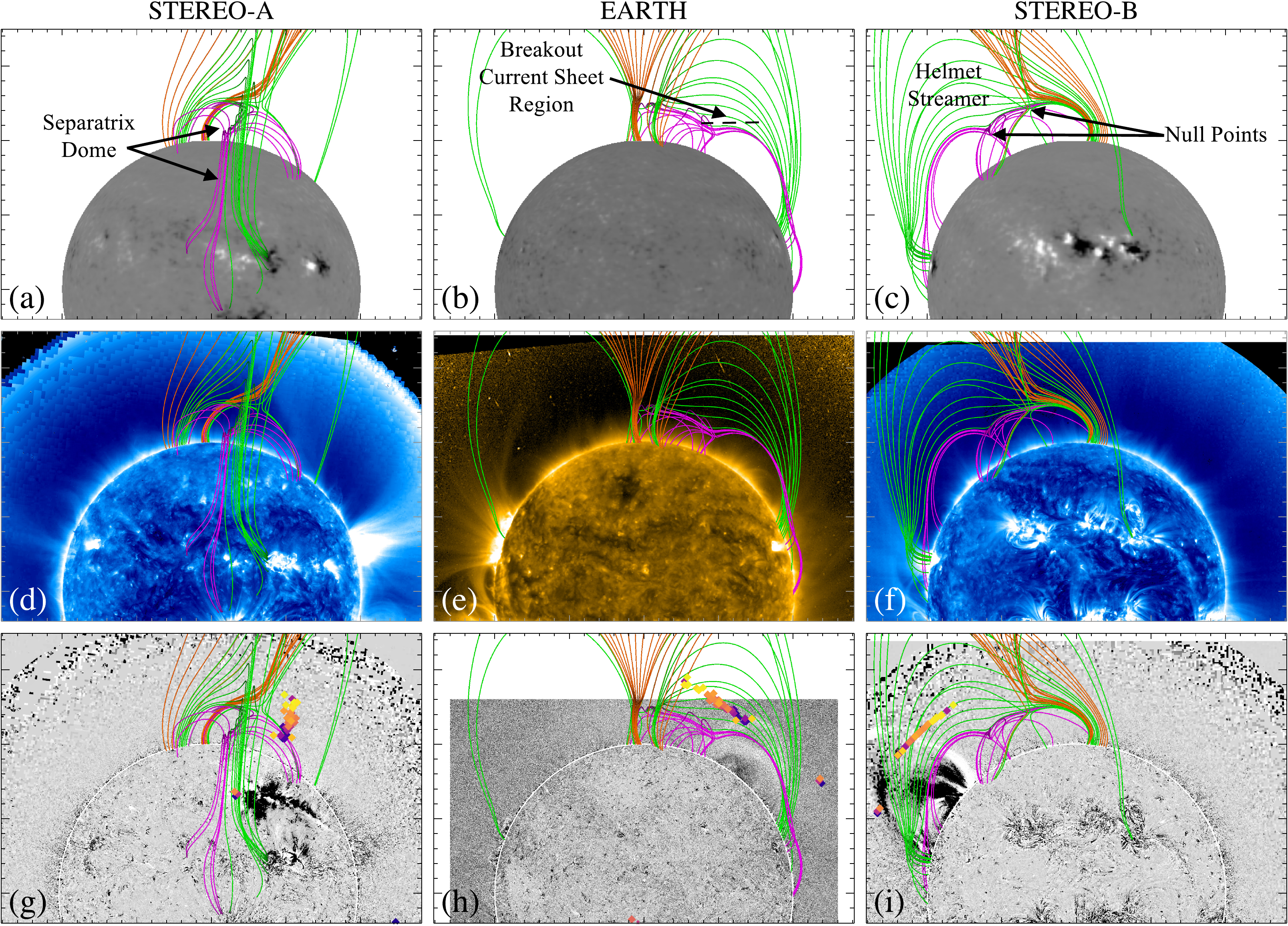}
      \caption{Pre-eruptive magnetic field configuration in relation to the CME eruption and radio emission. (a--c) PFSS field lines generated from the GONG synoptic magnetogram for CR 2126. (d--f) PFSS field lines plotted over the pre-eruptive configuration, as observed at 171~{\AA} by STEREO/SECCHI/EUVI and the Sun-Watcher with Active Pixel System and Image Processing \citep[SWAP;][]{se13} instrument onboard the Project for On Board Autonomy 2 \citep[PROBA2;][]{sa13} spacecraft. (g--i) PFSS field lines plotted over difference images of the CME eruption, as observed by STEREO/SECCHI/EUVI and SDO/AIA, overlaid with the radio centroids from Figs.~\ref{fig3}a and \ref{fig4}e-f.}
         \label{fig5}
   \end{figure*}

{Models of the coronal magnetic field are a valuable tool for providing context and assisting with the interpretation of observed CME eruption dynamics. Figure~\ref{fig5} shows a series of representative magnetic field lines associated with the CME source region and adjacent flux systems derived from the Potential Field Source Surface \citep[PFSS; e.g.][]{wa92, sc03} extrapolation from the Carrington Rotation (CR) 2126 synoptic map based on magnetograph observations by the National Solar Observatory (NSO) Global Oscillation Network Group \citep[GONG;][]{ha96}. Figure 5 shows the viewing perspectives of STEREO-A, Earth, and STEREO-B. The top row shows the CR~2126 synoptic magnetogram data projected onto the solar disc for each of the three viewpoints. The middle and bottom rows show the same PFSS field lines plotted over the pre-eruptive configuration and the running-difference images during the eruption from Figs.~\ref{fig3} and \ref{fig4}. In each panel the field lines are colour-coded to indicate their global connectivity: the orange field lines are open, the green field lines belong to the closed helmet streamer belt flux systems, and the magenta field lines are drawn near the separatrix dome and the external spine of a multipolar flux system beneath the helmet streamer.}

{Further examination of the PFSS topological structure reveals a pair of null points connected via a magnetic separator \citep[e.g.][]{ti12} and the portion of the separatrix dome that extends up to the polar region is essentially coincident with the boundary of the helmet streamer belt and the open field regions indicating an extremely sharp gradient in magnetic connectivity \citep{ti07}. Energization of these flux systems will form current sheets along the separatrix and quasi-separatrix surfaces that can facilitate magnetic reconnection and the rapid reconfiguration of the stressed fields \citep{an99,au05,ly16}. Recent modelling efforts have demonstrated breakout eruption scenarios for both coronal jets \citep{wy19} and CMEs \citep{ma19} in precisely this type of magnetic configuration.} 

{The overlay of field lines in Fig.~\ref{fig5}g--i  clearly shows the erupting portion of the multipolar flux system and overlying helmet streamer belt. The eruption proceeds towards the north in the direction of the closest pre-eruptive null point and separator line. The non-radial propagation (deflection) of CMEs toward their overlying breakout current sheet is expected and has been demonstrated in numerical simulations \citep[e.g.][]{to11,ly13}. During the rapid CME acceleration associated with the impulsive phase of eruptive flares, the CME ejecta may become fast enough to drive a shock in the low corona \citep{liu08,po11}. The shock can propagate into the overlying (closed or open) flux system(s) and quickly become much larger than the ejecta itself \citep[see e.g.\ Fig.~3b of][]{ly13}. If the EUV wave is the low-coronal signature of the 3D leading edge bubble and CME-driven shock \citep{as09,do12}, then the shock front and overlying breakout current sheet would both map to the extent and location of the radio emission shown in Fig.~\ref{fig1}.}


\section{Discussion and conclusions} \label{sec:conclusions}

{There are a number of radio observations that have been interpreted as consistent with breakout CME evolution in the low corona---specifically breakout reconnection occurring above and at the flanks of the main eruption. \citet{ma03} showed off-limb type II radio bursts at similar frequencies and spatial locations to those analysed here. Studies by \citet{au11,au13} have also shown similar patterns off limb and on disc but at slightly higher frequencies (200--400~MHz). The type II radio emission observed with the NDA most likely originated due to the CME-driven shock, while some components of the extended radio burst observed by the NRH could also be generated by electrons accelerated in the breakout reconnection region. Both mechanisms could produce radio emission in the regions reported here, and therefore it is not possible to discern between the two in the case of the NRH emission at higher frequencies. Furthermore, the NRH observations could consist of a combination of different types of radio emission. It is possible that in the NRH images we observed both a type IV radio burst originating from electrons accelerated at the breakout current sheet along the separatrix surfaces, as well as possible harmonic lanes of a type II radio burst originating from electrons accelerated at the CME shock. } 

{Here we present the analysis of an extended radio emission source above the limb associated with a back-side solar eruption on 22 July 2012. Multispacecraft observations from STEREO provide the additional viewpoints necessary to allow identification of the CME source region and to follow the eruption dynamics through the low corona. These viewpoints enable the 3D reconstruction of the radio emission locations, meaning that we can determine their position along the radio line of sight that is often difficult in CME--type II observations. Combining the remote-sensing observations with PFSS modelling of the large-scale magnetic field of the source region reveals a complex multipolar topology favourable to breakout reconnection during the CME onset and evolution. The spatial location and timing of the extended radio emission is consistent with both the production of a CME-driven shock and breakout reconnection above the CME eruption during its propagation through the streamer belt flux system towards the open field region over the north pole.}


\begin{acknowledgements}{The results presented here have been achieved under the framework of the Finnish Centre of Excellence in Research of Sustainable Space (Academy of Finland grant number 312390), which we gratefully acknowledge. B.J.L. acknowledges NASA HSR NNX17AI28G, LWS 80NSSC19K0088, and NSF AGS-1851945. E.K.J.K. acknowledges the European Research Council (ERC) under the European Union's Horizon 2020 Research and Innovation Programme Project SolMAG 724391, and Academy of Finland Project 310445. We would like to acknowledge the Nan{\c c}ay Radioheliograph, funded by the French Ministry of Education and the R\'egion Centre in France, and the Nan{\c c}ay Decametric Array hosted at the  Nan{\c c}ay Radio Observatory/Unit\'e Scientifique de Nan{\c c}ay of the Observatoire de Paris (USR 704-CNRS, supported by Universit\'e de Orl\'eans, OSUC and the R\'egion Centre). We would also like to thank the SDO, STEREO, and ARTEMIS spectrograph teams for providing free access to their data.}\end{acknowledgements}


\bibliographystyle{aa} 
\bibliography{aanda_arxiv} 

\end{document}